\documentclass[twocolumn,showpacs,superscriptaddress,prb,amsmath,amssymb]{revtex4}
\usepackage{times}
\usepackage{amsmath,bm,amsfonts}
\usepackage{dcolumn}
\usepackage{graphicx}
\usepackage{latexsym}

\begin{document}

\title{Spin Triplet Excitations for a Valence Bond Solid on the Kagome Lattice}

\author{Bohm-Jung \surname{Yang}}

\affiliation{Department of Physics and Astronomy, Center for Strongly
Correlated Materials Research, and Center for Theoretical Physics,
Seoul National University, Seoul 151-747, Korea}

\author{Yong Baek \surname{Kim}}

\affiliation{Department of Physics, University of Toronto,
Toronto, Ontario M5S 1A7, Canada}
\affiliation{School of Physics,
Korea Institute for Advanced Study, Seoul 130-722, Korea}

\author{Jaejun \surname{Yu}}

\affiliation{Department of Physics and Astronomy, Center for
Strongly Correlated Materials Research, and Center for Theoretical
Physics, Seoul National University, Seoul 151-747, Korea}

\author{Kwon \surname{Park}}
\email{kpark@kias.re.kr}

\affiliation{School of Physics, Korea Institute for Advanced
Study, Seoul 130-722, Korea}

\date{\today }

\begin{abstract}
One of the most promising candidate ground states for the quantum
antiferromagnetic Heisenberg model on the Kagome lattice is the
valence bond solid (VBS) with a 36-site unit cell. We present a
theory of triplet excitation spectra about this ground state using
bond operator formalism. In particular we obtain dispersions of
all 18 triplet modes in the reduced Brillouin zone. In the bond
operator mean-field theory, it is found that a large number of
triplet modes are non-dispersive. In particular, the lowest
triplet excitation is non-dispersive and degenerate with a
dispersive mode at the zone center. Away from the zone center, the
lowest triplet is separated from two other flat modes by a small
energy gap. Quantum fluctuations are considered by taking into
account scattering processes of two triplets and their bound state
formation, which leads to a downward renormalization of the lowest
spin triplet gap. The dispersion of the lowest triplet excitation
in the VBS state is compared with the dispersive lower bound of
the triplet continuum expected in competing spin liquid phases.
Implications to future neutron scattering experiments are
discussed.
\end{abstract}

\pacs{74.20.Mn, 74.25.Dw}

\maketitle

\section{\label{sec:intro} Introduction}

The quantum antiferromagnetic Heisenberg model on the Kagome
lattice is a quintessential example of frustrated quantum magnets
and therefore has been a subject of intense research activities
\cite{Zeng,Marston,sachdev2,Huse_old,Leung,Mila,Sindzingre_old,Hastings,Nikolic,Ran,Huse,Huse2}.
The ground state of this model, however, has been highly
controversial generating a number of competing proposals. Recent
proposals include a gapless spin liquid with Dirac fermionic
spinons \cite{Hastings,Ran}, a gapped spin liquid with bosonic
spinons \cite{sachdev2}, and the valence bond solid (VBS) with a
36-site unit cell structure \cite{Marston,Huse,Huse2}.

The identification of the ground state may be important to explain
a series of recent experiments on herbertsmithites
ZnCu$_3$(OH)$_6$Cl$_{2}$
\cite{kagome_exp1,kagome_exp2,kagome_exp3,kagome_exp4,kagome_exp5},
where spin-1/2 moments reside on the ideal Kagome-lattice
structure and interact with each other antiferromagnetically. It
was found that the material remains paramagnetic down to 50 mK
while the Curie-Weiss temperature is about 300 K
\cite{kagome_exp1,kagome_exp2,kagome_exp3,kagome_exp4,kagome_exp5}.
While this discovery is consistent with nonmagnetic ground states,
the precise nature of the ground state still needs to be clarified
by future experiments.

Recent theoretical studies reveal that the energy difference
between various competing ground states is extremely small
\cite{Hastings,Nikolic,Ran,Huse,Huse2}, which makes it quite
difficult to theoretically determine the true ground state. At the
same time, this implies that even a small perturbation to the
ideal nearest-neighbor Heisenberg model on the Kagome lattice may
result in completely different ground states. Given that small
perturbations such as the Dzyaloshinski-Moriya interaction and
impurities are indeed likely to be present in herbertsmithites
\cite{rigol}, the identification of the true ground state in this
material becomes a complex issue. In this context, it is crucial
to make testable experimental predictions for various proposed
ground states, which then can be used to determine the true nature
of the nonmagnetic state discovered in herbertsmithites
\cite{kagome_exp1,kagome_exp2,kagome_exp3,kagome_exp4,kagome_exp5}.

\begin{figure}[t]
\centering
\includegraphics[width=6.5cm]{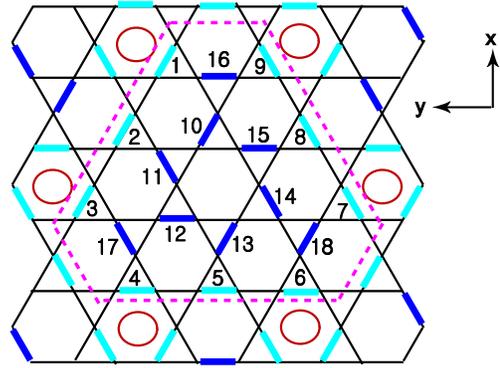}
\caption{(Color online) Dimer covering pattern for the valence
bond solid phase with a 36-site unit cell.
The unit
cell with 36 sites (or 18 dimers) is indicated by the enclosing
dotted line. Thick solid bars denote singlet dimers. We use
numbers between 1 and 9 to represent surrounding dimers (light-blue
bars) while core dimers (dark-blue bars) are denoted by numbers
between 10 and 18. Red circles indicate the locations of perfect
hexagons. Note that perfect hexagons are made of three
neighboring dimers, $(1, 4, 7)$ or $(3, 6, 9)$. The dimers 2, 5,
and 8 become bridges connecting perfect hexagons.}
 \label{fig:VBSpattern}
\end{figure}

In this paper, we focus on the valence bond solid ground state
with a 36-site unit cell (shown in Fig.~\ref{fig:VBSpattern}) and
compute spin-triplet excitation spectra by using bond opertor
formalism. The triplet excitation spectra can be directly measured
in future neutron scattering experiments when single crystals
become available. The eighteen dimers in a 36-site unit cell can
be categorized into two groups: ``core" dimers (numbered $10-18$
in Fig.~\ref{fig:VBSpattern}) including a ``pin-wheel" structure
at the center and ``surrounding" dimers (numbered $1-9$ in
Fig.~\ref{fig:VBSpattern}) including the honeycomb-lattice
structure of ``perfect hexagons". As shown later, triplet
excitations in the core dimers are highly localized and have flat
dispersions. On the other hand, triplet excitations in the
surrounding dimers can develop a dispersion by hopping around
perfect hexagons.


Triplet excitation spectra are obtained by using the fully
self-consistent mean-field theory in the bond operator
representation. As shown later (see Fig.\ref{fig:DispersionQuad}
and Fig.\ref{fig:DispersionQuartic}), the main feature of the triplet
excitation spectra is that a large number of triplet modes are
non-dispersive.
The lowest triplet excitation
is non-dispersive and degenerate with a dispersive mode at the
zone center. Away from the zone center, the lowest triplet is
separated from two other flat modes by a small energy gap.
It is expected that the triplet excitation energies at high symmetric
points such as $\Gamma$ and $X$ may be useful for comparison to
future neutron scattering experiments on single crystals.

The mean-field energy of the ground state is $-0.427 J$ which is
not too bad for a n\"{a}ive mean-field theory when compared to
$-0.438 J$ obtained from exact diagonalization of a 36-site
cluster \cite{Leung}. On the other hand, the lowest spin-triplet
gap is rather high, namely $0.795 J$, which is expected to be
significantly reduced once quantum fluctuations are properly taken
into account. It is shown later that the formation of two-triplet
bound state can renormalize the lowest spin-triplet gap from
$0.795 J$ to $0.622 J$. While it is still larger than $0.164 J$
obtained from exact diagonalization of a 36-site cluster
\cite{Sindzingre}, it is certainly in the right direction.
Possible origins for the discrepancy are discussed later.


The rest of the paper is organized as follows. In
section~\ref{sec:bondoperator}, physical motivations for studying
the VBS state with a 36-site unit cell are provided. In this
section, the bond operator representation is explained. In
section~\ref{sec:hamiltonian}, the bond operator mean-field theory
is developed. Results for the spin-triplet excitation spectra and
quantum fluctuation effects are presented in
section~\ref{sec:results}. Finally, in
section~\ref{sec:conclusion} we conclude by making a direct
comparison between the lowest spin-triplet excitation obtained
from our theory and those from the spin liquid theories. In the
spin liquid states, triplet excitations form a continuum. The
lower bound of such continuum, or the threshold energy, is given
by the convolution of spinon-antispinon excitation spectra.
Detailed measurement of the triplet spectra would be a key to the
identification of the true ground state.


\section{\label{sec:bondoperator} Motivation for Valence Bond Solid}

In this section we discuss physical motivations for the valence
bond solid (VBS) state with a 36-site unit cell structure. The VBS
state with a 36-site unit cell was initially proposed by Marston
and Zeng \cite{Marston} who envisioned the Kagome lattice as a
honeycomb-lattice arrangement of perfect hexagons composed of
resonant spin singlets.

Some time later, Nikolic and Senthil \cite{Nikolic} provided an
argument for the general validity of the 36-site VBS state in the
Kagome-lattice antiferromagnet by a duality mapping. The main
idea of this work is that the original Kagome-lattice
antiferromagnetic model can be mapped to the fully frustrated
Ising model on the dual dice lattice with transverse fields. Using
a reasonable assumption about the magnitude of the transverse fields
relative to the Ising coupling parameter, it is shown that the
honeycomb structure of perfect hexagons is likely to be the ground
state. Meanwhile, in a series expansion study by Singh and Huse
\cite{Huse}, it has been shown that the ground state energy is minimized
when perfect hexagons are connected to their neighbors through
empty triangles sharing a singlet dimer bond which in turn implies
the same honeycomb structure of perfect hexagons as discussed
in earlier works.

Here we provide an alternative viewpoint on the origin of the
valence bond solid state. We begin by asking what state may be the most
stable dimer covering configuration on the Kagome lattice. We
argue that the most stable configuration is the dimer covering
which maximizes the number of ``topologically perpendicular"
spin-singlet dimers. This argument is motivated by the exact
ground state of the antiferromagnetic Heisenberg model on the
Shastry-Sutherland lattice, where all dimers are
mutually perpendicular to each other \cite{ss1}. We use the word
``topologically perpendicular" since two neighboring dimers have
the identical spin exchange energy as those obtained in the
Shastry-Sutherland lattice when a spin belonging to a given dimer
shares a triangle with its neighboring dimer.

It can be shown that the pinwheel structure at the core of the
unit cell in Fig.~\ref{fig:VBSpattern} is the configuration
maximizing the number of ``topologically perpendicular" dimers.
The rest of the dimer covering falls into the honeycomb array of
perfect hexagons. While the final conclusion is exactly the same
as before, there is an additional advantage over the previous
arguments. Our point of view naturally leads to the fact that core
dimers are decoupled from surrounding dimers. To see this, it is
convenient to use a mathematical formalism which uses the
spin-singlet degree of freedom as a natural building block of the
VBS phase. The bond operator representation is such a formalism
\cite{sachdev,Gopalan,kpark}.

Let us consider two neighboring $S=\frac{1}{2}$ spins, ${\bf S}_R$
and ${\bf S}_L$. The Hilbert space is spanned by four states which
can be decomposed into a singlet state, $|s\rangle$, and three
triplet state, $|t_{x}\rangle$, $|t_{y}\rangle$ and
$|t_{z}\rangle$. Then, singlet and triplet boson operators are
introduced such that each of the above states can be created from
the vacuum $|0\rangle$ as follows:
\begin{align}
|s\rangle  &=s^{\dagger} |0\rangle  =\frac{1}{\sqrt{2}}
(|\uparrow\downarrow\rangle-|\downarrow\uparrow\rangle ),
\nonumber\\
|t_{x}\rangle  &=t_{x}^{\dagger} |0\rangle  =-\frac{1}{\sqrt{2}}
(|\uparrow\uparrow\rangle-|\downarrow\downarrow\rangle ),
\nonumber\\
|t_{y}\rangle  &=t_{y}^{\dagger} |0\rangle  =\frac{i}{\sqrt{2}}
(|\uparrow\uparrow\rangle+|\downarrow\downarrow\rangle ),
\nonumber\\
|t_{z}\rangle  &=t_{z}^{\dagger} |0\rangle  =\frac{1}{\sqrt{2}}
(|\uparrow\downarrow\rangle+|\downarrow\uparrow\rangle ).
\end{align}
To eliminate unphysical states from the enlarged Hilbert space,
the following constraint needs to be imposed on the bond-particle
Hilbert space:
\begin{equation} \label{eq:constraint}
s^{\dagger}s +t_{\alpha}^{\dagger}t_{\alpha} = 1,
\end{equation}
where $\alpha=x,y,$ and $z$, and we adopt the summation convention
for the repeated index hereafter unless mentioned otherwise.

Constrained by this equation,
the exact expressions for the spin operators can be written in
terms of the bond operators.
\begin{align}\label{eq:bond-ops}
S_{R\alpha}&=\frac{1}{2}(s^{\dag}t_{\alpha} +t_{\alpha}^{\dag}s
-i\varepsilon_{\alpha\beta\gamma}t_{\beta}^{\dag}t_{\gamma}),
\nonumber\\
S_{L\alpha}&=\frac{1}{2}(-s^{\dag}t_{\alpha} -t_{\alpha}^{\dag}s
-i\varepsilon_{\alpha\beta\gamma}t_{\beta}^{\dag}t_{\gamma}),
\end{align}
where $\varepsilon_{\alpha\beta\gamma}$ is the third rank
antisymmetric tensor with $\varepsilon_{xyz}=1$.

When neighboring dimers are ``topologically perpendicular",
spin-singlet contributions from both of the constituent spins of the
neighboring dimer exactly cancel out. This cancellation can be
seen in the bond operator representation of spin operators in
Eq.~(\ref{eq:bond-ops}) where it is shown that a pair of the spin
operators within the same dimer have the opposite sign in the
parts containing spin singlets. This results in the Hamiltonian
with no dispersive quadratic part for core triplets, which
eventually gives rise to nine-fold degenerate flat bands plotted
in Fig.~\ref{fig:DispersionQuad} and \ref{fig:DispersionQuartic}
in Sec.~\ref{sec:results}. This fact is completely consistent with
conclusions from the series expansion studies \cite{Huse2,Huse}.

In conclusion, the honeycomb lattice of perfect hexagons is a
stable dimer covering maximizing the number of ``topologically
perpendicular" dimers, which in turn naturally suggests decoupling
of core dimers from surrounding ones. Since the bond operator
representation is a perfect theoretical framework for such
situation, we use it to analyze the Kagome-lattice
antiferromagnetic Heisenberg model.

Since we expect qualitatively different behaviors between core and
surrounding dimers, we introduce a different set of parameters for
their singlet condensate densities, $\langle s_{{\bf i},n}\rangle$
= $\bar{s}_{n}$ and chemical potentials, $\mu_{{\bf i},n}$. Here ${\bf i}$ denotes the
position of the unit cell located at ${\bf i}$ and $n$ indicates the
dimer index inside the unit cell.
Furthermore, it is expected that the dynamics is different for
the dimers within perfect hexagons (the dimers 1, 3, 4, 6, 7, and 9 in
Fig.~\ref{fig:VBSpattern}) and those bridging perfect hexagons
(the dimers 2, 5, 8 in Fig.~\ref{fig:VBSpattern}). Therefore, we
introduce $\bar{s}_{C}$, $\mu_{C}$ for nine core dimers,
$\bar{s}_{H}$, $\mu_{H}$ for six surrounding dimers of
perfect hexagons, and $\bar{s}_{Br}$, $\mu_{Br}$ for three
bridging dimers connecting perfect hexagons. Technical details of
the bond operator analysis are provided in the next section.
Readers who are only interested in the results may directly go to
Sec.~\ref{sec:results}.

\section{\label{sec:hamiltonian} Hamiltonian and the Mean Field Theory}

We consider the following Hamiltonian for the Heisenberg model:
\begin{align}
H_J = \sum_{\langle {\bf r},{\bf r}' \rangle} J_{{\bf r},{\bf r}'}
{\bf S}({\bf r}) \cdot {\bf S}({\bf r}')
\end{align}
where ${\bf r}$ indicates the original coordinate of the lattice,
$J_{{\bf r},{\bf r}'}=J$ within dimers, and $J_{{\bf r},{\bf
r}'}=\lambda J$ between neighboring sites belonging to different
dimers. Utilizing the bond operator representation of spin
operators, the Hamiltonian can be rewritten solely in terms of bond
particle operators. At this point the hard-core constraint among
bond particle operators is imposed via the Lagrange multiplier
method;
\begin{align}
H_\mu = -\sum_{{\bf i},n} \mu_{{\bf i},n} (\bar{s}^2_{
n}+t^{\dagger}_{{\bf i},n\alpha}t_{{\bf i},n\alpha}-1),
\end{align}
where $({\bf i},n)$ denotes the position of the $n$-th dimer in
the unit cell located at ${\bf i}$. The chemical potential,
$\mu_{{\bf i},n}$, is set to be $\mu_C$ for core dimers, $\mu_H$
for perfect hexagon dimers, and $\mu_{Br}$ for bridge dimers.
Similarly, the spin-singlet condensate density, $\bar{s}_{n}$,
is set to be $\bar{s}_C$ for core dimers, $\bar{s}_H$ for
perfect hexagon dimers, and $\bar{s}_{Br}$ for bridge dimers.

The total Hamiltonian, $H=H_J+H_\mu$, can be written as
follows:
\begin{equation}
H=N\epsilon_{0} +H_\textrm{Quad, Core}+H_\textrm{Quad,
Surrounding}+H_\textrm{Quartic} , \label{H_total}
\end{equation}
where $H_\textrm{Quad, Core}$ denotes the quadratic part of the
Hamiltonian for core triplets while $H_\textrm{Quad, Surrounding}$
represents for surrounding triplets. In the above, $N$ is the number of
unit cells and
\begin{align}
\epsilon_{0}
&=9\left[\mu_{C}(1-{\bar{s}_{C}}^{2})-\frac{3}{4}J{\bar{s}_{C}}^{2}\right]\nonumber\\
&+6\left[\mu_{H}(1-{\bar{s}_{H}}^{2})-\frac{3}{4}J{\bar{s}_{H}}^{2}\right]\nonumber\\
&+3\left[\mu_{Br}(1-{\bar{s}_{Br}}^{2})-\frac{3}{4}J{\bar{s}_{Br}}^{2}\right].
\end{align}
It is important to note that the Hamiltonian does not
have the coupling between core
and surrounding triplets at the quadratic level. As shown in Sec.
\ref{sec:results}, the absence of quadratic coupling is
crucial for a complete decoupling between core
and surrounding triplets.

The quadratic Hamiltonian for core dimers is given by
\begin{align}\label{H_core} H_\textrm{Quad, Core}
&=\left(\frac{J}{4}-\mu_{C}\right)\sum_{\textbf{k}}\sum_{n=10}^{18}
t^{\dag}_{n\alpha}(\textbf{k})t_{n\alpha}(\textbf{k}),
\end{align}
where $\alpha=x,y,$ and $z$.
Also, the quadratic Hamiltonian for
surrounding dimers is given by
\begin{align}
H_\textrm{Quad, Surrounding} &=
H_\textrm{Quad,0}+H_\textrm{Quad,A}
+H_\textrm{Quad,B}+H_\textrm{Quad,C} \label{H_surrounding}
\end{align}
where $H_\textrm{Quad,0}$ denotes the quadratic Hamiltonian of triplet
operators
within a unit cell. $H_\textrm{Quad,A}$, $H_\textrm{Quad,B}$, and
$H_\textrm{Quad,C}$ describe quadratic coupling between
a given unit cell and neighboring unit cells separated by displacement
vectors, ${\bf r}_A$, ${\bf r}_B$, and ${\bf r}_C$, respectively.
These displacement vectors are defined as follows:
\begin{align}
{\bf r}_{A}&=4\sqrt{3}a\hat{x},  \nonumber \\
{\bf r}_{B}&=2\sqrt{3}a\hat{x}-6a\hat{y}, \nonumber \\
{\bf r}_{C}&=\textbf{${\bf r}_{B}$}-\textbf{${\bf
r}_{A}$}=-2\sqrt{3}a\hat{x}-6a\hat{y} ,
\end{align}
where $a$ is the distance between nearest neighbor spins.
More explicitly, we get
\begin{align}
&H_\textrm{Quad,0}=\left(\frac{J}{4}-\mu_{H}\right)\sum_{\textbf{k}}\sum_{n
\in g_{H}} t^{\dag}_{n\alpha}(\textbf{k})t_{n\alpha}(\textbf{k})
\nonumber\\
&\qquad+\left(\frac{J}{4}-\mu_{Br}\right)\sum_{\textbf{k}}\sum_{n
\in g_{Br}} t^{\dag}_{n\alpha}(\textbf{k})t_{n\alpha}(\textbf{k})
\nonumber\\
&+\frac{{\bar{s}_{H}\bar{s}_{Br}}\lambda J}{4}\sum_{(m,n)\in
G_{s}}\sum_{\textbf{k}} \Big\{ [
t^{\dag}_{m\alpha}(\textbf{k})t_{n\alpha}(\textbf{k})+\textrm{H.
c.}]
\nonumber\\
&\qquad\qquad\qquad+[
t^{\dag}_{m\alpha}(\textbf{k})t^{\dag}_{n\alpha}(-\textbf{k})+\textrm{H.
c.}] \Big\}
\end{align}
with $g_{H} = \{1,4,7,3,6,9 \}$, $g_{Br} = \{2,5,8 \}$, and $
G_{S}=\{(1,2),(2,3),(4,5),(5,6),(7,8),(8,9)\}$. Furthermore,
\begin{align}
&H_\textrm{Quad,A}=
\nonumber\\
&-\frac{{\bar{s}_{H}}^{2}\lambda J}{4}\sum_{(m,n)\in
G_{A,H}}\sum_{\textbf{k}} \Big\{ [e^{i {\bf k}\cdot{\bf r}_{A}}
t^{\dag}_{m\alpha}(\textbf{k})t_{n\alpha}(\textbf{k})+\textrm{H.
c.}]
\nonumber\\
&\qquad\qquad\qquad+[ e^{i{\bf k}\cdot{\bf
r}_{A}}t^{\dag}_{m\alpha}(\textbf{k})t^{\dag}_{n\alpha}(-\textbf{k})+\textrm{H.
c.}]
\Big\} \nonumber\\
&-\frac{{\bar{s}_{H}\bar{s}_{Br}}\lambda J}{4}\sum_{(m,n)\in
G_{A,Br}}\sum_{\textbf{k}} \Big\{ [e^{i{\bf k}\cdot{\bf r}_{A}}
t^{\dag}_{m\alpha}(\textbf{k})t_{n\alpha}(\textbf{k})+\textrm{H.
c.}]
\nonumber\\
&\qquad\qquad\qquad+[ e^{i{\bf k}\cdot{\bf
r}_{A}}t^{\dag}_{m\alpha}(\textbf{k})t^{\dag}_{n\alpha}(-\textbf{k})+\textrm{H.
c.}] \Big\},
\end{align}
where
\begin{align}
G_{A,H}=\{(1,4),(9,6)\} ,\quad   G_{A,Br}=\{(1,5),(9,5)\}.
\end{align}
$H_\textrm{Quad,B}$ and $H_\textrm{Quad,C}$ can be
obtained from $H_\textrm{Quad,A}$ by replacing (i) ${\bf r}_A$ by ${\bf
r}_B$ and ${\bf r}_C$ and (ii) $(G_{A,H}, G_{A,Br})$ by $(G_{B,H},
G_{B,Br})$ and $(G_{C,H}, G_{C,Br})$, respectively. Here,
\begin{align}
G_{B,H}&=\{(9,3),(7,4)  \},\quad   G_{B,Br}=\{(8,3),(8,4)  \}, \nonumber\\
G_{C,H}&=\{(7,1),(6,3)  \},\quad   G_{C,Br}=\{(7,2),(6,2) \}.
\end{align}

The quartic interaction part of the Hamiltonian is analyzed in the
Hartree-Fock mean-field theory via quadratic decoupling. The
resulting mean-field Hamiltonian is written as follows:
\begin{align}
H_\textrm{Quartic} &=H_\textrm{Quartic,0}+H_\textrm{Quartic,A}
+H_\textrm{Quartic,B}+H_\textrm{Quartic,C}
\end{align}
where, similar to the quadratic counterparts,
$H_\textrm{Quartic,0}$ denotes the quartic Hamiltonian of triplet
operators within a
unit cell. $H_\textrm{Quartic,A}$, $H_\textrm{Quartic,B}$, and
$H_\textrm{Quartic,C}$ describe quartic coupling between a given
unit cell and neighboring unit cells separated by displacement
vectors, ${\bf r}_A$, ${\bf r}_B$, and ${\bf r}_C$, respectively.
More explicitly, we obtain
\begin{align}
&H_\textrm{Quartic,0}=\frac{9}{2}\lambda J(Q_{C}^{2}-P_{C}^{2}+Q_{CS}^{2}-P_{CS}^{2})
\nonumber\\
&+\frac{3}{2}\lambda J(Q_{H}^{2}-P_{H}^{2})+\frac{3}{2}\lambda
J(Q_{Br1}^{2}-P_{Br1}^{2}+Q_{Br2}^{2}-P_{Br2}^{2})
\nonumber\\
&+\frac{\lambda J}{4}\sum_{(m,n)\in G_{S}}\sum_{\textbf{k}} \Big\{
[P_{Br1}
t^{\dag}_{m\alpha}(\textbf{k})t_{n\alpha}(\textbf{k})+\textrm{H.
c.}]
\nonumber\\
&\qquad\qquad\qquad-[Q_{Br1}
t^{\dag}_{m\alpha}(\textbf{k})t^{\dag}_{n\alpha}(-\textbf{k})+\textrm{H.
c.}]
\Big\} \nonumber\\
&+\frac{\lambda J}{2}\sum_{(m,n)\in G_{C}}\sum_{\textbf{k}} \Big\{
[P_{C}
t^{\dag}_{m\alpha}(\textbf{k})t_{n\alpha}(\textbf{k})+\textrm{H.
c.}]
\nonumber\\
&\qquad\qquad\qquad-[Q_{C}
t^{\dag}_{m\alpha}(\textbf{k})t^{\dag}_{n\alpha}(-\textbf{k})+\textrm{H.
c.}]
\Big\} \nonumber\\
&+\frac{\lambda J}{2}\sum_{(m,n)\in G_{CS}}\sum_{\textbf{k}}
\Big\{ [P_{CS}
t^{\dag}_{m\alpha}(\textbf{k})t_{n\alpha}(\textbf{k})+\textrm{H.
c.}]
\nonumber\\
&\qquad\qquad\qquad-[Q_{CS}
t^{\dag}_{m\alpha}(\textbf{k})t^{\dag}_{n\alpha}(-\textbf{k})+\textrm{H.
c.}] \Big\}
\end{align}
where
\begin{align}
G_{C}&=\{(10,11),(11,12),(12,13),(13,14),(14,15),\nonumber\\
&\qquad(15,10),(10,16),(12,17),(14,18)  \}, \nonumber\\
G_{CS}&=\{(16,1),(16,9),(17,4),(17,3),(18,7),(18,6),\nonumber\\
&\qquad(11,2),(13,5),(15,8) \}.
\end{align}
One also gets
\begin{align}
&H_\textrm{Quartic,A} \nonumber\\
&=\frac{\lambda
J}{4}\sum_{(m,n)\in G_{A,H}}\sum_{\textbf{k}} \Big\{ [P_{H}e^{i
{\bf k}\cdot{\bf r}_{A}}
t^{\dag}_{m\alpha}(\textbf{k})t_{n\alpha}(\textbf{k})+\textrm{H.
c.}]
\nonumber\\
&\qquad\qquad\qquad-[Q_{H} e^{i {\bf k}\cdot{\bf
r}_{A}}t^{\dag}_{m\alpha}(\textbf{k})t^{\dag}_{n\alpha}(-\textbf{k})+\textrm{H.
c.}]
\Big\} \nonumber\\
&+\frac{\lambda J}{4}\sum_{(m,n)\in G_{A,Br}}\sum_{\textbf{k}}
\Big\{ [P_{Br2}e^{i {\bf k}\cdot{\bf r}_{A}}
t^{\dag}_{m\alpha}(\textbf{k})t_{n\alpha}(\textbf{k})+\textrm{H.
c.}]
\nonumber\\
&\qquad\qquad\qquad-[Q_{Br2} e^{i {\bf k}\cdot{\bf
r}_{A}}t^{\dag}_{m\alpha}(\textbf{k})t^{\dag}_{n\alpha}(-\textbf{k})+\textrm{H.
c.}] \Big\},
\end{align}
Similar to the quadratic case, $H_\textrm{Quartic,B}$ and
$H_\textrm{Quartic,C}$ can be obtained from $H_\textrm{Quartic,A}$ by
replacing (i) ${\bf r}_A$ by ${\bf r}_B$ and ${\bf r}_C$ and (ii)
$(G_{A,H}, G_{A,Br})$ by $(G_{B,H}, G_{B,Br})$ and $(G_{C,H},
G_{C,Br})$, respectively.

The above mean-field order parameters, $P_{H}$, $Q_{H}$,
$P_{Br1}$, $Q_{Br1}$, $P_{Br2}$, $Q_{Br2}$, $P_{C}$, $Q_{C}$,
$P_{CS}$, and $Q_{CS}$, are determined by solving a coupled set of
ten self-consistency equations. In other words,
\begin{align}\label{eq:orderparameter}
P_{\gamma} &\equiv \langle t^{\dag}_{{\bf i}, n\alpha} t_{{\bf j},
m \alpha} \rangle, \;\; Q_{\gamma} \equiv \langle t_{{\bf i}, n
\alpha} t_{{\bf j}, m \alpha} \rangle
\nonumber\\
\end{align}
where $({\bf i},{\bf j})$ indicates the positions of the
neighboring unit cells and $\gamma \in (C, H, Br1, Br2, CS)$, both
of which are related to the dimer index pair, $(n,m)$. To be
specific, if $(n,m) \in G_C$, $\gamma=C$ and ${\bf j}={\bf i}$. If
$(n,m) \in G_S$, $\gamma=Br1$ and ${\bf j}={\bf i}$. On the other
hand, if $(n,m) \in G_{A,H}$, $\gamma=H$ and ${\bf j}-{\bf i}=
{\bf r}_A$. Those for $G_{B,H}$ and $G_{C,H}$ are defined
similarly. Also, if $(n,m) \in G_{A,Br}$, $\gamma=Br2$ and ${\bf
j}-{\bf i}= {\bf r}_A$. Those for $G_{B,Br}$ and $G_{C,Br}$ are
defined similarly. Finally, if $(n,m) \in G_{CS}$, $\gamma=CS$ and
${\bf j}={\bf i}$.

\begin{figure}[t]
\centering
\includegraphics[width=8cm]{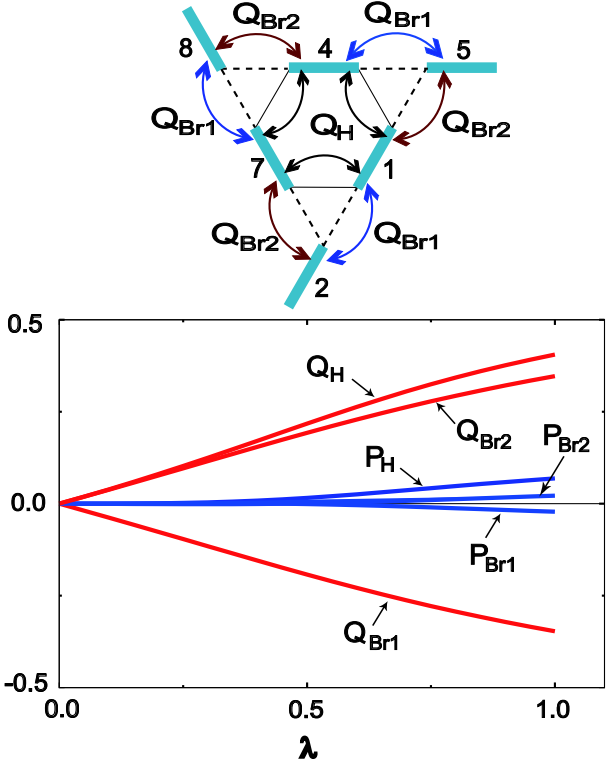}
\caption{(Color online) Mean-field order parameters, $P$ and $Q$,
as a function of $\lambda$. The top figure depicts dimer covering
of the valence bond solid state in the vicinity of a perfect
hexagon, which also shows how various $Q$'s are defined. Note that
each $Q$ describes an off-diagonal correlation between two dimers
connected by corresponding arrows. Diagonal correlations, $P_{H}$,
$P_{Br1}$, and $P_{Br2}$, are defined in the same way. It is
interesting to note that $Q_{Br1}$ $(P_{Br1})$ and $Q_{Br2}$
$(P_{Br2})$ have opposite signs. } \label{fig:OrderParameter}
\end{figure}

In physical terms, $P_{C}$ and $Q_{C}$ describe the diagonal and
off-diagonal triplet correlations between core dimers, respectively.
In the case
of surrounding dimers, six order parameters are introduced. First,
$P_{H}$ and $Q_{H}$ denote correlations between dimers within
perfect hexagons. The other four order parameters, $(P_{Br1},
Q_{Br1})$ and $(P_{Br2}, Q_{Br2})$, represent correlations between
a bridge dimer and two neighboring dimers belonging to the nearby
perfect hexagons. In particular, $(P_{Br1}, Q_{Br1})$ describes
the correlations between a given bridge dimer and the dimer lying
inside the nearby perfect hexagon in its parallel direction.
$(P_{Br2}, Q_{Br2})$ denotes the other correlations. A schematic
diagram showing the definition of these order parameters is
provided in the top panel of Fig.\ref{fig:OrderParameter}.
Finally, $P_{CS}$ and $Q_{CS}$ indicate the correlations between
core and surrounding dimers.

In order to determine the ten mean-field order parameters defined above,
one needs to compute yet another six unknown parameters which are
three spin-singlet condensate densities, $\bar{s}_{C}$,
$\bar{s}_{H}$, and $\bar{s}_{Br}$, and three chemical potentials,
$\mu_{C}$, $\mu_{H}$, and $\mu_{Br}$. The hard-core constraints
for bond operators provide three equations:
\begin{align}
&\bar{s}_{C}^{2}=1-\frac{1}{9 N}\sum_{n=10}^{18}\sum_{\textbf{k}}
\langle
t^{\dag}_{n\alpha}(\textbf{k})t_{n\alpha}(\textbf{k})\rangle,
\nonumber\\
&\bar{s}_{H}^{2}=1-\frac{1}{6 N}\sum_{n \in
g_{H}}\sum_{\textbf{k}} \langle
t^{\dag}_{n\alpha}(\textbf{k})t_{n\alpha}(\textbf{k})\rangle,
\nonumber\\
&\bar{s}_{Br}^{2}=1-\frac{1}{3 N}\sum_{n \in
g_{Br}}\sum_{\textbf{k}} \langle
t^{\dag}_{n\alpha}(\textbf{k})t_{n\alpha}(\textbf{k})\rangle.
\end{align}
On the other hand, the energy minimization with respect to
spin-singlet condensate densities generate the other three
necessary equations:
\begin{align}
\frac{\partial \varepsilon_{gr}}{\partial \bar{s}_{C}}=0, \quad
\frac{\partial \varepsilon_{gr}}{\partial \bar{s}_{H}}=0, \quad
\frac{\partial \varepsilon_{gr}}{\partial \bar{s}_{Br}}=0.
\end{align}

Now the ground state energy and excitation spectra can be
obtained as follows. For this
purpose, the mean-field Hamiltonian is rewritten in the following
compact fashion:
\begin{align}
H=\frac{1}{2}\sum_{\textbf{k}} {\bf \Lambda}_{\alpha}^{\dag} {\bf M}
{\bf \Lambda}_{\alpha} -\frac{3}{4} \textrm{Tr} {\bf M},
\end{align}
where
\begin{align}
{\bf \Lambda}_{\alpha}^{\dagger}(\textbf{k}) &\equiv \left[ t^{\dag}_{1
\alpha}(\textbf{k}), \ldots, t^{\dag}_{18 \alpha}(\textbf{k}),
t_{1 \alpha}(-\textbf{k}), \ldots, t_{18 \alpha}(-\textbf{k})
\right]
\end{align}
and the matrix, ${\bf M}$, is determined by simply reorganizing
the mean-field Hamiltonian.

It is important to note that our Hamiltonian describes dynamics of
boson operators, in which case obtaining normal modes of the
Hamiltonian is not equivalent to diagonalizing the matrix ${\bf
M}$ in the above. Instead, we need to consider the following
eigenvalue problem \cite{Blaizot}:
\begin{align}
{\bf I}_{B}{\bf M} {\bf \Psi} = \omega {\bf \Psi}
\end{align}
where
\begin{align}
{\bf I}_{B}= \left( \begin{array}{ccc}
{\bf I} & {\bf 0} \\
{\bf 0} & -{\bf I} \\
\end{array} \right)
\end{align}
with ${\bf I}$ being the $18 \times 18$ identity matrix and
\begin{align} \label{eq:eigenvector}
{\bf \Psi} = \left( \begin{array}{ccc}
{\bf \eta} \\
{\bf \xi} \\
\end{array} \right)
\end{align}
with ${\bf \eta}^t=(\eta_{1},\ldots ,\eta_{18})$ and ${\bf
\xi}^t=(\xi_{1},\ldots ,\xi_{18})$. This difference between
fermionic and bosonic problems fundamentally originates from the
difference in their operator commutation relations.

\section{\label{sec:results} Results}

\subsection{Ground state energy}

The ground state energy per unit cell is obtained by solving the
eigenvalue equation described in the preceding section. In other
words,
\begin{align}
\varepsilon_{gr} &=\epsilon_{0}+\frac{9}{2}\lambda J
(Q_{C}^{2}-P_{C}^{2}+Q_{CS}^{2}-P_{CS}^{2})
\nonumber\\
&\qquad+\frac{3}{2}\lambda J (Q_{H}^{2}-P_{H}^{2})
\nonumber \\
&\qquad+\frac{3}{2} \lambda J
(Q_{Br1}^{2}-P_{Br1}^{2}+Q_{Br2}^{2}-P_{Br2}^{2})
\nonumber \\
&\qquad+\frac{3}{2}\frac{1}{N}\sum_{\textbf{k}}\sum_{n=1}^{18}
\omega_{n}(\textbf{k})-\frac{18}{2}\Big(\frac{J}{4}-\mu_{H}\Big)
\nonumber \\
&\qquad-\frac{9}{2}\Big(\frac{J}{4}-\mu_{Br}\Big)
-\frac{27}{2}\Big(\frac{J}{4}-\mu_{C}\Big),
\end{align}
where $\omega_n$ $(n=1,\dots,18)$ are triplet eigenenergies
obtained from diagonalization. As mentioned previously, the ground
state energy is minimized with respect to spin-singlet condensate
densities. Note that the minimization process is performed
simultaneously satisfying the ten self-consistency conditions and
three hard-core constraints.

Now let us discuss numerical results. When quartic interactions
are completely ignored, the ground state energy per site is found
to be $-0.414 J$ at $\lambda = 1$. The inclusion of
quartic interactions lowers the ground state energy to $-0.427 J$
per site, which can be favorably compared with $-0.438 J$ from exact
diagonalization \cite{Leung}. This result is quite encouraging given that
it is obtained from a naive mean-field theory. This perhaps indicates
the robustness of the VBS state.

\subsection{Triplet dispersions without quartic interactions}

We now discuss the energy dispersion of triplet excitations. In
this section quartic interaction terms are ignored. The analysis
of quartic interaction effects is relegated to the next section.
One of the reasons why we first focus on the quadratic part of the
Hamiltonian is that the overall structure of the dispersion is
well captured even at the quadratic level. The inclusion of
quartic interaction terms modifies the dispersion only
quantitatively.

\begin{figure}[t]
\centering
\includegraphics[width=6.5cm]{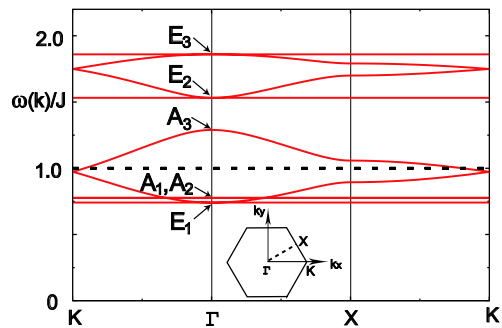}
\caption{(Color online) Energy dispersion of the triplet
excitations when quartic interaction effects are ignored. The
dotted line describes the nine-fold degenerate triplet states from
core dimers. The solid lines denote the nine triplet states from
surrounding dimers. } \label{fig:DispersionQuad}
\end{figure}

As seen in Eq.~(\ref{H_core}), core triplets have no dispersions.
In this situation the ground-state energy minimization condition
immediately leads to the conclusion that $\bar{s}_{C} = 1$ and
$\mu_{C} = -0.75 J$, which implies that in the core part of the unit
cell triplet fluctuations are entirely absent and triplet
excitations are completely localized. As mentioned previously, the
reason for this complete localization has to do with the dimer
covering structure of the valence bond solid phase. The key fact
is that every core dimer is ``topologically perpendicular" to its
neighboring dimers in the sense that each spin belonging to a core
dimer shares a triangle with its neighboring dimer. This results
in the Hamiltonian with no dispersive quadratic part for core
triplets, which gives rise to nine-fold degenerate flat bands
plotted as a dotted line in Fig.~\ref{fig:DispersionQuad}.

Now let us turn to the remaining nine triplets in the surrounding
dimers. After diagonalizing the corresponding bosonic Hamiltonian
via the method described in the preceding section, mean-field
parameters, $\bar{s}_{H}$,$\bar{s}_{Br}$, $\mu_{H}$, and
$\mu_{Br}$, are self-consistently determined.
Fig.~\ref{fig:DispersionQuad} shows the resulting dispersion of
nine triplet excitations at the isotropic exchange limit of
$\lambda=1$. It is convenient at this point to classify
characteristics of these eigenmodes using group theory. Since
the unit cell possesses the $C_{3}$ point group symmetry, the
states at the $\Gamma$ point can be decomposed with respect to
irreducible representations of the $C_{3}$ point group. A
representation, $R$, can be decomposed in the basis containing
nine triplets localized at each dimer as follows:
\begin{align}\label{eq:group}
R=3 A \bigoplus 3 E
\end{align}
where $A$ is an one-dimensional representation and $E$ is a
two-dimensional one. (See Ref.~\onlinecite{Tinkham} for standard
conventions and notations.)

\begin{figure}[t]
\centering
\includegraphics[width=7cm]{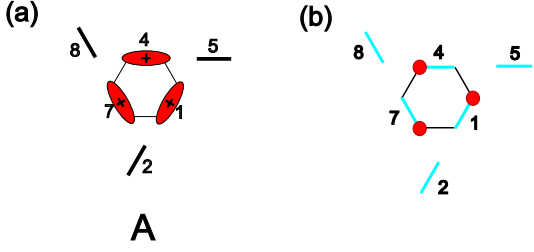}
\caption{(Color online) Real space representation of the
eigenvectors of the flat A mode at the $\Gamma$ point. The size of
ovals and the sign inside indicate relative weights and phases of
the amplitude at each dimer, respectively. (a) One of the flat $A$
modes with the energy eigenvalue $\Omega_{A}$. Triplets
around each perfect hexagon have the same phases and weights.
Two different flat $A$ modes are distinguished by relative phases
between neighboring perfect hexagons. (b) Local geometry around a
perfect hexagon. Here red dots indicate the position of R-spins
within a dimer. See Eq.~(\ref{eq:bond-ops}) for the definition of
the R-spin.} \label{fig:Amode}
\end{figure}

\begin{figure}[t]
\centering
\includegraphics[height=7cm]{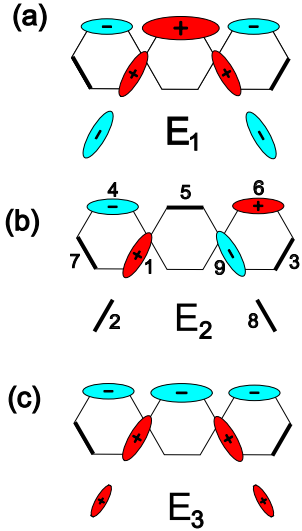}
\caption{(Color online) Real space representation of the
eigenvectors of the flat E modes. To avoid degeneracy eigenvectors
are computed slightly away from the $\Gamma$ point to the
direction of the K point, for (a) $E_1$ mode, (b) $E_2$ mode, and (c)
$E_3$ mode. Note that the size of ovals and the sign
indicate relative weights and phases of the amplitude at each
dimer, respectively. } \label{fig:Emode}
\end{figure}

The lowest flat mode belonging to the E irreducible representation
is degenerate with a dispersive mode at the $\Gamma$ point ($E_{1}$ mode
in Fig.\ref{fig:DispersionQuad}). There
are two other flat modes ($A_{1}$ and $A_{2}$) which are separated from the lowest flat
mode by a small gap. These states have $A$ characters and are
completely localized within perfect hexagons. The fifth dispersive
mode ($A_{3}$) is another $A$ mode. The remaining four modes belong to the E
irreducible representation and make two doubly degenerate states
at the $\Gamma$ point ($E_{2}$ and $E_{3}$). Real space representations of the five flat
mode eigenvectors are shown in Fig.~\ref{fig:Amode} and
\ref{fig:Emode}. Details of these eigenvectors for non-dispersive modes
are further analyzed in the next section where their
explicit forms are also presented. In the next section
we study quartic interaction effects via the self-consistent
mean-field theory.


\subsection{Triplet dispersions with quartic interactions}

In this section we solve the total Hamiltonian containing
contributions from all eighteen dimers in the unit cell. The fully
self-consistent calculation, however, shows that core dimers are
still completely decoupled from surrounding dimers and remain flat
in energy. In other words,
\begin{align}
P_{C}=Q_{C}=P_{CS}=Q_{CS}=0, \quad \bar{s}_{C}=1, \quad
\mu_{C}=-0.75 J.
\end{align}

\begin{figure}[t]
\centering
\includegraphics[width=7cm]{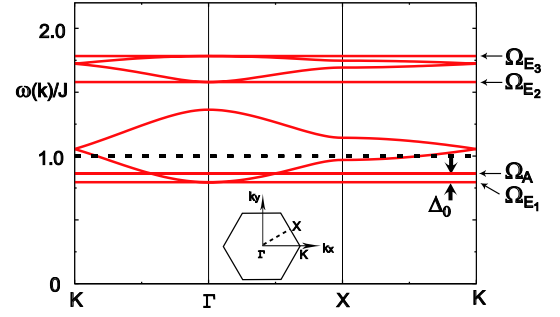}
\caption{(Color online) Energy dispersion of the triplet
excitations when quartic interaction effects are included in the
self-consistent mean-field theory. Away from the zone center,
the lowest flat mode is
separated from doubly-degenerate flat $A$ modes by a small gap,
$\Delta_{0}$ $\cong$ 0.069 J. Eigenenergies of the flat modes are
shown on the right hand side of the plot, where subscripts
indicate the irreducible representation of each state at the
$\Gamma$ point. } \label{fig:DispersionQuartic}
\end{figure}

On the other hand, surrounding dimers experience band
renormalization due to finite triplet correlations, $P$ and $Q$.
Order parameters for the triplet correlations are plotted as a
function of $\lambda$ in Fig.\ref{fig:OrderParameter}. The order
parameters are shown to have the largest value inside perfect
hexagons which in turn implies that vacuum fluctuation
is strongest at this location. Triplet correlations between bridge
and perfect hexagon dimers are also quite large, about $86 \%$ of
those between perfect hexagon dimers. It is interesting to note
that $Q_{Br1}$ $(P_{Br1})$ and $Q_{Br2}$ $(P_{Br2})$ are the same
in magnitude but are the opposite in sign. In
Fig.\ref{fig:DispersionQuartic} we plot the triplet dispersions
obtained from fully self-consistent parameters at $\lambda=1$. The
overall structure is identical to that obtained by ignoring
quartic interaction terms.

One of the most interesting characteristics of the triplet
dispersion is the existence of a large number of flat bands. Among
the nine modes coming from surrounding dimers, five states have no
dispersion. (If one includes the flat modes originating from core
dimers, fourteen states out of eighteen are flat in energy.) These
five flat modes are categorized in terms of the $C_3$ point group
symmetry at the $\Gamma$ point. In the right hand side of
Fig.\ref{fig:DispersionQuartic}, each mode is indicated by
$\Omega_{A_1}$, $\Omega_{A_2}$, $\Omega_{E_{1}}$,
$\Omega_{E_{2}}$, and $\Omega_{E_{3}}$, respectively
(these are also energy eigenvalues). Note that
$\Omega_{A}$ = $\Omega_{A_{1}}$ = $\Omega_{A_{2}}$ and
$\Omega_{E_{1}}$ $<$ $\Omega_{E_{2}}$ $<$ $\Omega_{E_{3}}$. In the
above notations subscripts denote irreducible representations to
which each flat mode belongs.

It is interesting to note that the eigenvalue equation can be
solved exactly for the flat modes. The precise analytic
expressions for the flat mode eigenvalues are given as follows:
\begin{align}\label{eq:eigenvalues}
\Omega_{A_{1}} &=\Omega_{A_{2}} =\sqrt{(a_{H} - 2 b_{H})^{2} - 4 q_{H}^{2}}, \nonumber\\
\Omega_{E_{2}} &=\sqrt{(a_{H} + b_{H})^{2} - q_{H}^{2}}, \nonumber\\
\Omega_{E_{1}} &=\Omega_{-}, \quad \Omega_{E_{3}} =\Omega_{+},
\end{align}
where
\begin{align}
\Omega_{\pm}^{2} =\frac{1}{2} \left[
a_{Br}^{2}+(a_{H}+b_{H})^{2}-q_{H}^{2}-12(q_{Br}^{2}-b_{Br}^{2})\pm
\sqrt{D} \right],
\end{align}
\begin{align}
D &=  \left[ a_{Br}^{2}-(a_{H}+b_{H})^{2}+q_{H}^{2}\right]^{2} \nonumber\\
&-24 (q_{Br}^{2}-b_{Br}^{2}) \left[ a_{Br}^{2}+(a_{H}+b_{H})^{2}-q_{H}^{2} \right]\nonumber\\
&+24 a_{Br}\Big[(q_{Br}+b_{Br})^{2}(a_{H}+b_{H}-q_{H}) \nonumber\\
&+(q_{Br}-b_{Br})^{2}(a_{H}+b_{H}+q_{H})\Big],
\end{align}
and
\begin{align}
a_{H}&=\frac{J}{4}-\mu_{H}, \quad a_{Br}=\frac{J}{4}-\mu_{Br}, \nonumber\\
b_{H}&=\frac{\lambda J}{4}(\bar{s}_{H}^{2}-P_{H}),\quad b_{Br}=\frac{\lambda J}{4}(\bar{s}_{H}\bar{s}_{Br}-P_{Br2}), \nonumber\\
q_{H}&=\frac{\lambda J}{4}(\bar{s}_{H}^{2}+Q_{H}),\quad
q_{Br}=\frac{\lambda J}{4}(\bar{s}_{H}\bar{s}_{Br}+Q_{Br2}).
\end{align}

It is important to notice that the eigenvalues, $\Omega_{A_{1}}$,
$\Omega_{A_{2}}$, and $\Omega_{E_{2}}$, are completely determined
by the parameters defined inside perfect hexagons. In fact, the
same energy eigenvalues are obtained by applying the bond operator
theory to a single isolated perfect hexagon using the mean-field
order parameters, $\mu_{H}$, $\bar{s}_{H}$, $P_{H}$, and $Q_{H}$.
This means that for the flat $A$ and $E_2$ modes a simple product
of perfect-hexagon eigenstates becomes that of the full Kagome
lattice, at least in the self-consistent mean-field theory.

As mentioned in the previous section, the flat $A$ modes are
completely localized within perfect hexagons, {\it i.~e.~}, there is no
weight for the bridge dimers. Here triplets around each perfect hexagon
have the same weight and phase while the $A_1$ and $A_2$ modes are
distinguished by a relative phase difference in neighboring
perfect hexagons. The $E_2$ mode is also localized within perfect
hexagons. The difference between the flat $A$ and $E_2$ modes
originates from the fact that they represent different eigenstates of
the isolated perfect-hexagon. Schematic diagrams for the real
space representation of these flat modes are provided in
Fig.~\ref{fig:Amode} and \ref{fig:Emode}, which show the relative
weight and sign of the eigenvector amplitudes at each real space
dimer location.

To understand the flat nature of the $A$ modes, one needs to
examine the local geometry around a perfect hexagon which is
plotted in Fig.~\ref{fig:Amode}(b). As seen in
Eq.~(\ref{eq:bond-ops}), two constituent spins, the $R$ and $L$
spins, within the same dimer are distinguished by the sign
difference in the part containing spin singlet operators. This is
basically due to the odd parity of the spin singlet under the
inversion with respect to the center of the dimer. The $R$-spins
are denoted by red dots in Fig.~\ref{fig:Amode}(b). Because of the
three-fold rotational symmetry, every spin from bridge dimers is
simultaneously connected to both $R$ and $L$ spins of the
perfect hexagon dimers. Due to the above-mentioned sign
difference, hopping amplitudes between perfect hexagon and bridge
dimers cancel, leading to the complete localization within perfect
hexagons. The flat $E_2$ mode can be understood similarly.

We now present precise analytic expressions for the eigenvectors
of all five flat modes in Table~\ref{table:fulleigenvector}. The
prefactors, $\tau_M$, in Table~\ref{table:fulleigenvector} denote
the relative magnitudes of the hole (particle) component in the
Bogoliubov quasiparticles (quasiholes), which are given as
follows:
\begin{align}
\tau_{A_{1}} &= \tau_{A_{2}} =\frac{a_{H} - 2 b_{H} -
\Omega_{H}}{2 q_{H}}, \nonumber
\end{align}
\begin{align}
\tau_{E_{2}} &= \frac{\Omega_{E_{2}}-a_{H} -
b_{H}}{q_{H}},\nonumber
\end{align}
\begin{align}
\tau_{E_{1}} &= \frac{(\Omega_{E_{1}}^{2}-a_{Br}^{2})(\Omega_{E_{1}}-a_{H}-b_{H})}{q_{H}(\Omega_{E_{1}}^{2}-a_{Br}^{2}) + 12 a_{Br}b_{Br}q_{Br}}\nonumber\\
&-\frac{  \Big\{6 \Omega_{E_{1}}(b_{Br}^{2}-q_{Br}^{2})+ 6
a_{Br}(b_{Br}^{2}+q_{Br}^{2})  \Big \} }
{q_{H}(\Omega_{E_{1}}^{2}-a_{Br}^{2}) + 12 a_{Br}b_{Br}q_{Br}},
\nonumber
\end{align}
\begin{align}
\tau_{E_{3}} &= \frac{(\Omega_{E_{3}}^{2}-a_{Br}^{2})(\Omega_{E_{3}}-a_{H}-b_{H})}{q_{H}(\Omega_{E_{3}}^{2}-a_{Br}^{2}) + 12 a_{Br}b_{Br}q_{Br}}\nonumber\\
&-\frac{  \Big\{6 \Omega_{E_{3}}(b_{Br}^{2}-q_{Br}^{2})+ 6
a_{Br}(b_{Br}^{2}+q_{Br}^{2})  \Big \} }
{q_{H}(\Omega_{E_{3}}^{2}-a_{Br}^{2}) + 12 a_{Br}b_{Br}q_{Br}}.
\end{align}
Note that $\tau_{A_1}$, $\tau_{A_2}$ and $\tau_{E_2}$ depend only
on the perfect hexagon parameters. The coefficients, $c_{1}$
$(d_1)$ and $c_{3}$ $(d_{3})$ indicate the relative magnitude of
coupling between perfect hexagon and bridge dimers in the $E_{1}$
and $E_{3}$ modes, respectively. More explicitly, one obtains
\begin{align}
c_{1} &=\frac{\Omega_{E_{1}}-a_{Br}}{b_{Br}+\tau_{E_{1}}q_{Br}},
\quad
d_{1}=\frac{-\Omega_{E_{1}}-a_{Br}}{b_{Br}+q_{Br}/\tau_{E_{1}}}, \nonumber\\
c_{3} &=\frac{\Omega_{E_{3}}-a_{Br}}{b_{Br}+\tau_{E_{3}}q_{Br}},
\quad
d_{3}=\frac{-\Omega_{E_{3}}-a_{Br}}{b_{Br}+q_{Br}/\tau_{E_{3}}}.
\end{align}
Note that the eigenvector amplitudes in bridge dimers, {\it i.~e.~},
$(\eta_2,\eta_5,\eta_8)$ and $(\xi_2,\xi_5,\xi_8)$, are zero for
the flat $A_1$, $A_2$ and $E_2$ modes, which confirms that they
are completely localized within perfect hexagons.

\begin{table}
\begin{tabular}{@{}|c|c|c|c|c|c|}
\hline \hline
& $A_{1}$ & $A_{2}$ & $E_{1}$ & $E_{2}$ & $E_{3}$  \\
\hline \hline
$\eta_{1}$ & $z_{1}^{*}z_{2}$ & $z_{1}^{*}z_{2}$ & $2-z_{1}^{*}-z_{1}^{*}z_{2}$ & $2-z_{1}^{*}-z_{1}^{*}z_{2}$ & $2-z_{1}^{*}-z_{1}^{*}z_{2}$ \\
$\eta_{4}$ & $z_{2}$          & $z_{2}$          & $2-z_{1}-z_{2}$              & $2-z_{1}-z_{2}$ & $2-z_{1}-z_{2}$ \\
$\eta_{7}$ & $ 1 $            & $1$              & $2-z_{2}^{*}-z_{2}^{*}z_{1}$ & $2-z_{2}^{*}-z_{2}^{*}z_{1}$ & $2-z_{2}^{*}-z_{2}^{*}z_{1}$ \\
\hline
$\eta_{2}$ & $ 0 $            & $0$              & $c_{1}(1-z_{2}z_{1}^{*})$    & $0$ & $c_{3}(1-z_{2}z_{1}^{*})$ \\
$\eta_{5}$ & $ 0 $            & $0$              & $c_{1}(1-z_{1})$             & $0$ & $c_{3}(1-z_{1})$ \\
$\eta_{8}$ & $ 0 $            & $0$              & $c_{1}(1-z_{2}^{*})$         & $0$ & $c_{3}(1-z_{2}^{*})$ \\
\hline
$\eta_{3}$ & $r_{1}z_{2}$          & $r_{2}z_{2}$         & $2-z_{2}-z_{1}^{*}z_{2}$     & $z_{2}+z_{1}^{*}z_{2}-2$ & $2-z_{2}-z_{1}^{*}z_{2}$ \\
$\eta_{6}$ & $r_{1}z_{1}$          & $r_{2}z_{1}$         & $2-z_{1}-z_{2}^{*}z_{1}$     & $z_{1}+z_{2}^{*}z_{1}-2$ & $2-z_{1}-z_{2}^{*}z_{1}$ \\
$\eta_{9}$ & $r_{1} $            & $r_{2}$             & $2-z_{1}^{*}-z_{2}^{*}$      & $z_{1}^{*}+z_{2}^{*}-2$ & $2-z_{1}^{*}-z_{2}^{*}$ \\
\hline \hline
$\xi_{1}/\tau_{M}$ & $z_{1}^{*}z_{2}$ & $z_{1}^{*}z_{2}$ & $2-z_{1}^{*}-z_{1}^{*}z_{2}$ & $2-z_{1}^{*}-z_{1}^{*}z_{2}$ & $2-z_{1}^{*}-z_{1}^{*}z_{2}$ \\
$\xi_{4}/\tau_{M}$ & $z_{2}$          & $z_{2}$          & $2-z_{1}-z_{2}$              & $2-z_{1}-z_{2}$ & $2-z_{1}-z_{2}$ \\
$\xi_{7}/\tau_{M}$ & $ 1 $            & $1$              & $2-z_{2}^{*}-z_{2}^{*}z_{1}$ & $2-z_{2}^{*}-z_{2}^{*}z_{1}$ & $2-z_{2}^{*}-z_{2}^{*}z_{1}$ \\
\hline
$\xi_{2}/\tau_{M}$ & $ 0 $            & $0$              & $d_{1}(1-z_{2}z_{1}^{*})$    & $0$ & $d_{3}(1-z_{2}z_{1}^{*})$ \\
$\xi_{5}/\tau_{M}$ & $ 0 $            & $0$              & $d_{1}(1-z_{1})$             & $0$ & $d_{3}(1-z_{1})$ \\
$\xi_{8}/\tau_{M}$ & $ 0 $            & $0$              & $d_{1}(1-z_{2}^{*})$         & $0$ & $d_{3}(1-z_{2}^{*})$ \\
\hline
$\xi_{3}/\tau_{M}$ & $r_{1}z_{2}$        & $r_{2}z_{2}$        & $2-z_{2}-z_{1}^{*}z_{2}$     & $z_{2}+z_{1}^{*}z_{2}-2$ & $2-z_{2}-z_{1}^{*}z_{2}$ \\
$\xi_{6}/\tau_{M}$ & $r_{1}z_{1}$        & $r_{2}z_{1}$        & $2-z_{1}-z_{2}^{*}z_{1}$     & $z_{1}+z_{2}^{*}z_{1}-2$ & $2-z_{1}-z_{2}^{*}z_{1}$ \\
$\xi_{9}/\tau_{M}$ & $r_{1} $            & $r_{2}$             & $2-z_{1}^{*}-z_{2}^{*}$      & $z_{1}^{*}+z_{2}^{*}-2$ & $2-z_{1}^{*}-z_{2}^{*}$ \\
\hline \hline
\end{tabular}
\caption{Eigenvectors of the five flat modes.
Among the 36 components of the full
eigenvector, ${\bf \Psi}^t =
(\eta_1,\ldots,\eta_{18},\xi_1,\ldots,\xi_{18})$, only the
components for surrounding dimers are shown. Note that core
dimers are decoupled from surrounding ones. In the above,
$z_1=\exp{(-i{\bf k}\cdot{\bf r}_A)}$, $z_2=\exp{(-i{\bf
k}\cdot{\bf r}_B)}$, and $(r_1,r_2)$ represents any pair of two complex
numbers satisfying $r_{1}^{*}r_{2} = -1$. The prefactor, $\tau_M$,
depends on the mode index, $M \in A_1, A_2, E_1, E_2,$ and $E_3$.
Explicit expressions for $\tau_M$ as well as $c_1,
c_3, d_1,$ and $d_3$ are provided in the text. }
\label{table:fulleigenvector}
\end{table}

While the amplitudes in bridge dimers do not vanish, the nature of
the $E_{1}$ and $E_{3}$ modes is rather similar to that of the
$E_2$ mode. As one can see from Table~\ref{table:fulleigenvector},
the amplitudes inside perfect hexagons for the $E_1$ and $E_3$
modes are precisely identical to those for the $E_2$ mode. This
identity is fundamentally due to (i) the odd parity of the singlet
operator and (ii) the three-fold rotational symmetry of the
lattice. Coupling to bridge dimers splits triple degeneracy by
lowering the energy for the $E_1$ mode and increasing it for the
$E_3$ mode. The difference from the $E_2$ mode case is that
hopping amplitudes between perfect hexagon and bridge dimers do
not cancel, but instead they add.

It is interesting to note that our $E_1$ mode
is fully consistent with the lowest triplet excitation
obtained in the series expansion study by Singh and Huse \cite{Huse2}.
In fact, after properly redefining the unit cell convention,
it can be shown that our $E_1$ mode becomes precisely equivalent
to the lowest energy eigenstate obtained in first order of $\lambda$
under the conditions that (i) all off-diagonal coupling terms are ignored, (ii) all
quartic interaction terms are ignored, and (iii) perfect hexagon
and bridge dimers are physically identical,
i.~e., $\bar{s}_H = \bar{s}_{Br}$ and $\mu_H = \mu_{Br}$.
Considering that our $E_1$ mode does not change abruptly upon
relaxing these conditions or restoring the full self-consistency,
one may expect that
our $E_1$ mode is indeed adiabatically connected to
the lowest energy eigenvector obtained in the previous
series expansion study \cite{Huse2} and perhaps
the corresponding eigenvector in
yet-to-be-studied higher order series expansion.

\subsection{Quantum fluctuations}
\label{sec:fluctuations}

Up to now, triplet interactions are treated within the
self-consistent mean-field theory. To investigate effects
of quantum fluctuations, we need to go beyond
the mean field theory.
There are three classes of order parameters in the mean-field
theory, which can be affected by quantum fluctuations. These
are chemical potentials, $\mu$, diagonal correlation parameters,
$P$, and off-diagonal correlation parameters, $Q$.
Spin-singlet condensate densities, $\bar{s}$,
which are the remaining variational parameters,
can be determined by minimizing the ground state energy
after quantum fluctuation effects are incorporated for the
above-mentioned order parameters. Below we check how
each of these order parameters is affected by quantum fluctuations.

The chemical potential is the Lagrange multiplier for the hard-core
constraint. In the usual weak-coupling limit where the singlet nature of the
ground state is robust, the dominant contribution comes from the
hard-core constraint, Eq.~(\ref{eq:constraint}), which can be
conveniently implemented by an infinite on-site repulsion
between triplets \cite{Kotov1}. When the triplet density is low,
this hard-core constraint can be treated by summing ladder diagrams
for the scattering vertex. The complete ladder diagram summation for the
full lattice is rather complicated. Fortunately, however, in our system
the most important aspect of the lowest excitation is determined by
the nature of eigenstates inside a single perfect hexagon. Therefore, we
expect that the dynamics inside a single perfect hexagon is a good indicator
for the full lattice as far as the lowest excitation is concerned.
It is shown that there is little difference between the chemical potential
obtained from the mean-field theory and that from the ladder diagram summation
using a single perfect hexagon. For this reason, we ignore the effects of
quantum fluctuations on the chemical potential.

Next, we consider the effects of quantum fluctuations on the
diagonal correlation order parameters, $P$, and the off-diagonal
correlation order parameters, $Q$. From our mean-field analysis,
it is shown that the shear size of the diagonal correlation
parameters is almost one order of magnitude smaller than that of the
off-diagonal counterparts (See Fig.~\ref{fig:OrderParameter}). It
is thus expected that, while $P$ is to be certainly renormalized
by quantum fluctuations, the overall size of its renormalization
should be much smaller than that of $Q$. Thus we
focus on $Q$ below while ignoring fluctuation effects on $P$.

Off-diagonal correlations are enhanced when there are strong
fluctuations toward the formation of two-triplet bound states,
which is caused by the attractive interaction between nearest
neighbor triplets \cite{Kotov2, Kotov3}.
Effects of these fluctuations on the singlet/triplet spectrum
can be captured by considering successive particle-particle
scattering processes which renormalize the triplet pair
emission/absorption amplitudes, {\it i.~e.~}, the coefficients of
$t^{\dagger}t^{\dagger}+\textrm{H. c.}$ terms \cite{Kotov4}.
To be concrete we describe below how the corrections in these
coefficients are computed.


In the self-consistent mean field calculation
as described in the Sec.~\ref{sec:hamiltonian}, the bare pair
emission/absoprtion amplitude, $B^0$, is renormalized to
$B^{MF}$ where
\begin{align}
B^0_\gamma = \Big\{ \begin{array}{lll}
\bar{s}^2_H & \textrm{if $\gamma = H$},\\
\bar{s}_H \bar{s}_{Br} & \textrm{if $\gamma = Br1$ or $Br2$}, \\
\end{array}
\end{align}
and $B^{MF}_\gamma = B^0_\gamma +Q_\gamma$.
To go beyond the mean-field theory, we consider scattering of two
triplets: $\alpha + \beta \rightarrow \mu + \nu$ where triplet
spin indices, $\alpha, \beta, \mu,$ and $\nu$, belong to $\{ x, y, z \}$.
From the quartic interaction terms, one can get the following bare
scattering amplitude:
\begin{align}
V_{\alpha \beta, \mu \nu}=-\frac{J}{4}(\delta_{\alpha
\beta}\delta_{\mu \nu}-\delta_{\alpha \nu}\delta_{\beta
\mu}),
\end{align}
which shows that
in the singlet $S=0$ channel the scattering amplitude is given
by $V^{(S=0)}=\frac{1}{3}\delta_{\alpha \beta}\delta_{\mu
\nu}V_{\alpha \beta, \mu \nu}=-\frac{J}{2}$.

\begin{figure}[t]
\centering
\includegraphics[width=7cm]{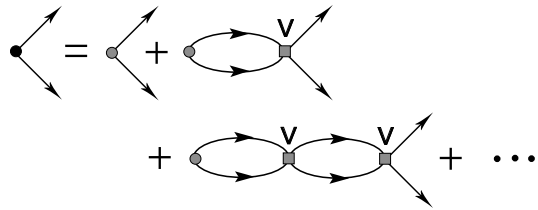}
\caption{Renormalization of the triplet pair emission (or
absorption) amplitude by quantum fluctuations. Vertices come from
the singlet $S=0$ scattering channels of two triplets. }
\label{fig:fluctuation}
\end{figure}

We now evaluate the ladder series for scattering processes in
Fig.~\ref{fig:fluctuation}, which renormalize $B^{MF}_\gamma$
as follows:
\begin{align}
B^{MF}_\gamma \quad \rightarrow \quad B_\gamma=B^{MF}_\gamma+\Delta B_\gamma,
\end{align}
where
\begin{align} \label{eq:fluctuation}
\Delta B_\gamma &= -\frac{B^{MF}_\gamma V^{(S=0)} \Pi_\gamma}{1+V^{(S=0)}\Pi_\gamma},
\end{align}
and
\begin{align}
\Pi_\gamma &= \sum_{\bf k}\int\frac{d\omega}{2 \pi i} G^{MF}_{nn}({\bf k},\omega)G^{MF}_{mm}(-{\bf k},-\omega).
\end{align}
In the above $G^{MF}_{nn}$ and $G^{MF}_{mm}$ are the mean-field
Green's function for triplets in the $n$- and $m$-th dimer
location in the unit cell. Conventions for the relationship
between $\gamma$ and $(n,m)$ are the same as those for $P$ and $Q$
in Eq.~(\ref{eq:orderparameter}).

The inclusion of the above quantum fluctuation corrections
modifies the triplet Hamiltonian matrix which, after
diagonalization, leads to a reduction of the lowest spin gap from
$0.795 J$ to $0.622 J$. Our prediction for the lowest spin gap is
still larger than $0.164 J$ obtained from exact diagonalization of
a 36-site cluster ~\cite{Sindzingre} or $0.08 \pm 0.02 J$ from a
recent 7th order series expansion result (Note that the value from
series expansion becomes $0.2 J$ in the same finite 36-site
cluster) \cite{Huse2}. This may suggest that quantum fluctuations
beyond what we have considered may be necessary to reach
quantitative agreement. At the same time, finite-size effects also
need to be examined very carefully.
%

\section{\label{sec:conclusion} Discussion}

The Kagome-lattice antiferromagnetic Heisenberg model has been
generally regarded as one of the most geometrically frustrated spin systems
in two dimension. Because of this, the Kagome lattice has been
considered as a promising candidate system for realizing exotic quantum spin
liquid ground states. In particular, there are two spin liquid
states that have received much attention lately: (i) the U(1)
Dirac spin liquid state suggested in a projected wave function
study by Ran {\it et al.} \cite{Ran} and (ii) the $Z_2$ spin
liquid state obtained in a bosonic large-$N$ Sp$(N)$ approach by
Sachdev \cite{sachdev2}.

\begin{figure}[t]
\centering
\includegraphics[width=8cm]{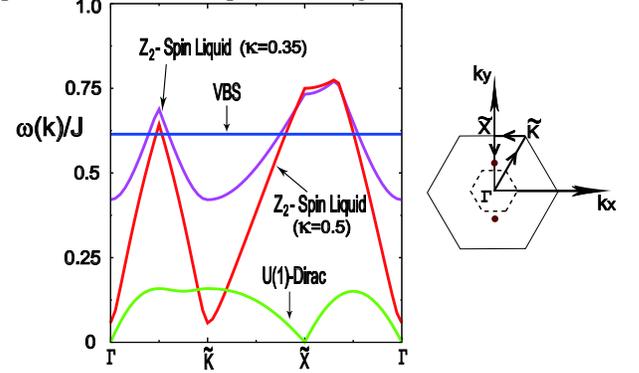}
\caption{(Color online) Spin-triplet excitation spectra for three
candidate ground states of the Kagome-lattice Heisenberg
antiferromagnet: (i) the valence bond solid (VBS) with a 36-site
unit cell, (ii) the U(1) Dirac spin liquid and (iii) the $Z_{2}$
spin liquid. In the case of the VBS state we plot the renormalized
spectrum of the lowest triplet mode including quantum fluctuation
effects as described in Sec.~\ref{sec:fluctuations}. For these two
spin liquid states we plot the lower bound of the two-spinon
continuum. The first Brillouin zone for the spin liquid states
which has a 3-site triangular unit cell is depicted on the right
hand side. The small dotted hexagon shows the first Brillouin zone
for the VBS state with a 36-site unit cell. Two red dots lying
in the $y$-axis indicate the positions (gauge-dependent) of
the Dirac nodes for the U(1) Dirac spin liquid state.} \label{fig:Comparison}
\end{figure}
To explicitly compare the lowest spin-triplet excitation energy
of the VBS state with those of the above-mentioned spin liquid states, we compute
the lower bound of the two-spinon continuum in spin liquid phases.
The lower bound (or the bottom of the continuum) is given by
\begin{align}
\omega({\bf k}) = \textrm{Min}_{\bf p} [ \epsilon({\bf k}/2+{\bf
p}/2) +\epsilon({\bf k}/2-{\bf p}/2) ],
\end{align}
where $\epsilon({\bf q})$ is the single spinon excitation energy
with momentum ${\bf q}$. An assumption behind this calculation is
that spinons themselves are weakly interacting.

For the U(1) Dirac spin liquid state we use the tight-binding
spinon Hamiltonian with the mean-field order parameter for spinon
hopping $\chi=0.221$ \cite{Hastings,Ran}. In the case of the
$Z_{2}$ spin liquid state we solve the full saddle point equations
for $\kappa=0.35$ and $\kappa=0.5$. Here $\kappa$ is a
parameter measuring the strength of quantum fluctuations. In the
large-$N$ Sp$(N)$theory the Kagome-lattice system shows magnetic
ordering when $\kappa$ is larger than $0.53$ \cite{sachdev2}.
Results for the lower bound of the two-spinon continuum are
plotted in Fig.~\ref{fig:Comparison}.

Owing to gapless fermionic spinons at nodal points, the U(1)-Dirac
spin liquid state exhibits gapless spin-triplet excitations at the
zero momentum point, $\Gamma$, and those momenta connecting two
Dirac nodes, $\widetilde{X}$. The lower bound of the two-spinon
continuum has variations in energy approximately given by $\chi
J$. Gapped bosonic spinons of the $Z_{2}$ spin liquid state make
the two-spinon spectrum also gapped with relatively large
variations for the lower bound. On the other hand, the lowest
spin-triplet excitation is completely non-dispersive in our
valence bond solid theory. Since the flat dispersion of the lowest
spin-triplet excitation is a distinct characteristic of the
valence bond solid state, it may be used to distinguish our
predictions from those of other spin liquid scenarios.

\begin{center}
\textbf{Acknowledgments}
\end{center}

This work was supported by the NSERC, CRC, CIAR,
KRF-2005-070-C00044 (YBK) and by the KOSEF through CSCMR SRC (BJY
and JY). YBK thanks Brad Marston for careful explanation of his past works
and acknowledges the Aspen Center for Physics where some part of this work
was initiated. BJY thanks to Choong H. Kim for his valuable comments
regarding numerics. Also, two of the authors (BJY, KP) would like
to thank Asia Pacific Center for Theoretical Physics (APCTP) for
its hospitality.





\begin{thebibliography}{99}

\bibitem{Zeng} C. Zeng and V. Elser, Phys. Rev. B {\bf 42}, 8436
(1990).

\bibitem{Marston} J. B. Marston and C. Zeng, J. Appl. Phys. {\bf
69}, 5962 (1991).

\bibitem{sachdev2} S. Sachdev, Phys. Rev. B {\bf 45}, 12377 (1992).

\bibitem{Huse_old} R. R. P. Singh and D. A. Huse, Phys. Rev.
Lett. {\bf 68}, 1766 (1992).

\bibitem{Leung} P. W. Leung and V. Elser, Phys. Rev. B {\bf 47},
5459 (1993).

\bibitem{Mila} F. Mila, Phys. Rev. Lett. {\bf 81}, 2356 (1998).

\bibitem{Sindzingre_old} P. Sindzingre, G. Misguich, C. Lhuillier,
B. Bernu, L. Pierre, Ch. Waldtmann and H.-U. Everts , Phys. Rev.
Lett. {\bf 84}, 2953 (2000).

\bibitem{Hastings} M. B. Hastings, Phys. Rev. B {\bf 63}, 014413 (2000).

\bibitem{Nikolic} P. Nikolic and T. Senthil, Phys. Rev. B
{\bf 68}, 214415 (2003).

\bibitem{Ran} Y. Ran, M. Hermele, P. A. Lee, and X. G. Wen, Phys. Rev. Lett.
{\bf 98}, 117205 (2007).


\bibitem{Huse}
R. R. P. Singh and D. A. Huse, Phys. Rev. B {\bf 76}, 180407(R)
(2007).

\bibitem{Huse2} R. R. P. Singh and D. A. Huse, arXiv:0801.2735.

\bibitem{kagome_exp1} J. S. Helton, K. Matan, M. P. Shores, E. A. Nytko,
B. M. Bartlett, Y. Yoshida, Y. Takano, A. Suslov, Y. Qiu, J.-H.
Chung, D. G. Nocera, and Y. S. Lee, Phys. Rev. Lett. {\bf 98},
107204 (2007).


\bibitem{kagome_exp2} P. Mendels, F. Bert, M. A. de Vries, A. Olariu, A. Harrison, F.
Duc, J. C. Trombe, J. S. Lord, A. Amato, and C. Baines , Phys.
Rev. Lett. {\bf 98}, 077204 (2007);

\bibitem{kagome_exp3} O. Ofer, A. Keren, E. A. Nytko, M. P. Shores, B. M. Bartlett, D.
G. Nocera, C. Baines, and A. Amato, cond-mat/0610540.

\bibitem{kagome_exp4} T. Imai, E. A. Nytko, B. M. Bartlett, M. P. Shores, and
D. G. Nocera, cond-mat/0703141.

\bibitem{kagome_exp5} M. A. de Vries, K. V. Kamenev, W. A. Kockelmann, J.
Sanchez-Benitez, and A. Harrison, arXiv:0705.0654.

\bibitem{rigol} M. Rigol and R. R. P. Singh, Phys. Rev. B {\bf 76}, 184403 (2007);
G. Misguich and P. Sindzingre, Eur. Phys. J. B {\bf 59}, 305 (2007).


\bibitem{Sindzingre} Unpublished work by P. Sindzingre and C. Lhuillier
as quoted by Singh and Huse \cite{Huse2}.

\bibitem{ss1} B. S. Shastry and B. Sutherland, Physica B {\bf 108}, 1069
(1981).

\bibitem{sachdev} S. Sachdev and R. N. Bhatt, Phys. Rev. B {\bf 41},
9323 (1990).

\bibitem{Gopalan} S. Gopalan, T. M. Rice and M. Sigrist, Phys. Rev. B
{\bf 49}, 8901 (1994).

\bibitem{kpark} K. Park and S. Sachdev, Phys. Rev. B
{\bf 64}, 184510 (2001).




\bibitem{Blaizot}
J. P. Blaizot and G. Ripka, Quantum Theory of Finite Systems, (The
MIT Press, 1986).

\bibitem{Tinkham}
M. Tinkham, Group Theory and Quantum Mechancies, (McGraw-Hill Book
Company, 1964).






\bibitem{Kotov1} V. N. Kotov, O. Sushkov, Zheng Weihong, and J. Oitmaa, Phys. Rev.
Lett. {\bf 80}, 5790 (1998).

\bibitem{Kotov2} O. P. Sushkov and V. N. Kotov, Phys. Rev. Lett. {\bf 81},
1941 (1998).

\bibitem{Kotov3} V. N. Kotov, O. P. Sushkov, and R. Eder, Phys. Rev. B. {\bf 59},
6266 (1999).

\bibitem{Kotov4} V. N. Kotov, D. X. Yao, A. H. Castro Neto, and D. K. Campbell,
arXiv:0704.0114.



\end{thebibliography}
\end{document}